\documentclass[twocolumn,pra,aps,showpacs]{revtex4}

\usepackage{psfrag,graphicx}

\usepackage{dcolumn}
\usepackage{amsmath,amssymb}
\usepackage{bm}
\usepackage{pstricks}

\def\half{\textstyle\frac{1}{2}}

\begin{document}



\title[Short Title]{
DMRG study of the Bond Alternating \textbf{S}=1/2 Heisenberg
ladder with Ferro-Antiferromagnetic couplings}

\author{
J. Almeida$^{\star}$, M.A. Martin-Delgado$^{\star}$ and G.
Sierra$^{\ast}$
 }
\affiliation{ $^{\star}$Departamento de F\'{\i}sica Te\'orica I,
Universidad Complutense. 28040 Madrid, Spain.
\\
$^{\ast}$Instituto de F\'{\i}sica Te\'orica, C.S.I.C.- U.A.M.,
Madrid, Spain. }

\begin{abstract}
We obtain the phase diagram in the parameter space $(J'/J, \gamma)$ and an accurate estimate
of the critical line separating the different phases. We show several measuments of the
magnetization, dimerization, nearest neighbours correlation, and density of energy in
the different zones of the phase diagram, as well as a measurement of the string order parameter
proposed as the non vanishing phase order parameter characterizing Haldane phases. All these
results will be compared in the limit $J'/J\gg 1$ with the behaviour of the $\textbf{S}=1$
Bond Alternated Heisenberg Chain (BAHC).
The analysis of our data supports the existence of a dimer phase
separated by a critical line from a Haldane one, which has exactly the same nature as the Haldane
 phase in the $\textbf{S}=1$ BAHC.
\end{abstract}


\maketitle


\section{Introduction}
\label{sect_intro}

Quantum systems when placed in low dimensional lattices typically exhibit strongly correlated
effects driving them towards regimes with no classical analog. Many properties of these regimes or quantum phases
\cite{sachdev_book99} depend in turn on the properties of their ground state and
low lying energy excitations \cite{jaitisi}.

A problem of particular interest in the field of strongly correlated systems is the emergence of critical
phases in a system where the generic behaviour as coupling constants are varied is to be a gapped system,
although those gapped phases may be of different nature. In this paper we address this problem by selecting
a system of quantum spins that allows us to perform a detailed study of critical and non-critical phases
on equal footing, i.e., without any bias towards an a priori  preferred phase. For reasons explained in
Sect.\ref{sect_model}, the quantum spins are arranged in a 2-leg ladder lattice \cite{dagotto_rice96}
with anti-ferromagnetic
Heisenberg couplings along the legs while rung couplings are ferromagnetic.
In addition, we also introduce an explicit dimerization coupling in the Hamiltonian along the leg
directions, which can be varied from zero to strong values. This coupling plays a major role in order to
create the aforementioned critical phases out of a system with only gapped phases.

This particular type of 2-leg ladder system has a number of
open problems such as the precise location of critical phases in the phase diagram of the coupling constants,
and the nature of the gapped phases it exhibits.
Our study is complete enough so as to be able to solve for these problems in a very precise manner.

The understanding of these purely quantum effects is usually a hard problem. Perturbative and variational methods
in quasi-one dimensional systems like chains and ladders are not well suited to uncover the physics in the
whole range of coupling constants involved in the description of the interactions in the system.
On the contrary the DMRG method \cite{white92}, \cite{white93}, \cite{hallberg06}, \cite{scholl05},
\cite{dmrg_book} allows us to identify the critical phases clearly and
without any bias. This is so because the method is non-perturbative and allows a controllable management of errors.

Our studies are also of interest since experiments on ladder materials have  revealed a very complex behaviour,
such as an interplay between
a spin-gapped normal states
and superconductivity \cite{dagotto}. Moreover, a new field of study for these complex effects has been
opened by the simulation of strongly correlated systems in optical lattices \cite{greiner}, in particular
quantum spin chains and ladders \cite{OLs}.

This paper is organized as follows:
 in Sect.\ref{sect_model} we introduce the model Hamiltonian \eqref{2legHam}
 describing a 2-leg ladder lattice of spins $S=\half$ with columnar bond-alternating
 antiferromagnetic couplings in the horizontal direction and ferromagnetic couplings in
 the vertical direction, see Fig.\ref{ladderpic}. We can identify some particular behaviours
 in appropriate weak and strong coupling limits, but not for generic values
 of the couplings. In Sect.\ref{sect_critical} we point out the rich physical effects posed
 by open boundary conditions in these 2-leg ladders with finite length, although it also implies
 an a priori analysis in order to find out which low-lying states contribute to the gap of the
 system in the thermodynamic limit. This we can be done with the DMRG method by targeting several states
 and measuring their magnetization properties in the bulk and at the ends. Then, we  compute
 numerically the gap and we establish the existence of a critical line in the quantum
 phase diagram of the model. A numerical fit of this critical curve is also given.
 In Sect.\ref{sect_haldanedimer} we determine  the structure of the phase diagram
 by identifying the type of gapped phases occurring at each side of the critical line found in the
 previous section. They correspond to Haldane and dimer phases. They are identified by measuring
 the string order parameter and the dimerization parameter with the DMRG method.
 We complete our study of these phases measuring different observables.
Sect.\ref{sect_conclusions} is devoted to conclusions.

\section{The model}
\label{sect_model}

\begin{figure}[tb]
\psfrag{XXXprime}{$J(1+\gamma)$}
\includegraphics[scale=0.7]{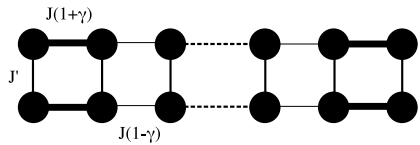}
\caption{Pictorial representation of the quantum Hamiltonian \eqref{2legHam}.
The geometry of the lattice is a 2-leg ladder.
Each solid dot is a spin $S=\half$. In the horizontal direction (legs), we picture
the bond alternation with strong links $J(1+\gamma)$ and weak links $J(1-\gamma)$.
In the vertical direction (rungs), the system is arranged in the form of a columnar
dimerization: strong links are parallel to one another, and similarly for weak links
in the lattice.}
\label{ladderpic}
\end{figure}

Competing ferromagnetic versus antiferromagnetic spin interactions
may give rise to critical phases if they are appropriately arranged
in certain quasi-one dimensional lattices. One emblematic example
of this phenomenon is a lattice of quantum spins with the shape
of a 2-leg ladder such that there are antiferromagnetic couplings along the
legs and ferromagnetic interactions along the rungs connecting both legs. In addition,
the antiferromagnetic couplings are bond-alternating in a columnar fashion.
Dimerization interactions in the Hamiltonian are also known as staggered
interactions. This configuration is shown in Fig.\ref{ladderpic}.
More precisely, this configuration of Heisenberg-like
interactions is associated with the following quantum Hamiltonian
\begin{equation}
\begin{split}
H= &J\sum_{\ell=1,2} \sum_{i=1}^{L-1}(1-(-1)^i\gamma)\mathbf{S}_{i}(\ell) \cdot
\mathbf{S}_{i+1}(\ell) \\
&+J'\sum_{i=1}^L \mathbf{S}_{i}(1) \cdot
\mathbf{S}_{i}(2),
\end{split}
\label{2legHam}
\end{equation}
where $\mathbf{S}_i(\ell)$ are quantum spin $S=\half$ operators located
at site $i$ of the leg $\ell$,
and $J>0$, $J'<0$, $\gamma \in [-1,1]$ are the antiferromagnetic,
ferromagnetic and staggering couplings, respectively, as mentioned above.

Notice that several known regimes can be reached by tuning the coupling
constants towards particular values.
In the weak coupling limit, making $\vert J'/J\vert\ll 1$
we end up with a system consisting on two effectively decoupled
$S=1/2$ Heisenberg chains with bond alternation (BAHC), which are known to be gapped
for every value of the dimerization parameter $\gamma$ \cite{hida92}, except for the point $\gamma=0$.
In the strong coupling limit, making $\vert J'/J\vert \gg 1, J'<0$ the system can be effectively
described by a $S=1$ spin chain with bond alternation, which is predicted to be gapped for all
values of $\gamma$ except for a critical point at a non-zero value $\gamma_c$ \cite{affleck_haldane87}.
These predictions are based on an approximate mapping onto the O(3) $\sigma$ model  \cite{haldane82}
at topological angle $\theta=2\pi S (1-\gamma)$. This yields a critical value of $\gamma_c=\half$
when $\theta=\pi$, and similarly another symmetric critical value at $\gamma_c=-\half$. Thus, we shall
always concentrate in the region $\gamma \geq 0$, due to the symmetry $\gamma \leftrightarrow -\gamma$
in the Hamiltonian \eqref{2legHam}. This non-linear sigma model (NL$\sigma$M) prediction
misses the correct location of the critical point due to the approximations involved in that mapping. 
The exact location of this point has been widely studied \cite{totsuka_et_al95} and results slightly varied depending on the approach, however modern studies place it at $\gamma_c=0.259$ 
\cite{Salerno2006}\cite{Kohno98}, also compatible with Fig.\ref{L500vsL2x140}(\textit{lower})
 which gives $\gamma_c=0.2590\pm0.0001$ for a chain of $500$ sites. These studies also conclude that the region $\vert\gamma\vert<\gamma_c$ corresponds to a 
Haldane phase while for $\vert \gamma\vert >\gamma_c$ we move to a dimer phase. 
The emergence of a dimerized $S=1$ spin chain in the strong coupling limit can be explained
by noting that as the rung coupling is ferromagnetic and strong $J'<0, |J'|\gg J$, the two spins $S=\half$
in each rung find energetically favorable to form a spin triplet.

\begin{figure}[h]
\includegraphics[scale=1.0]{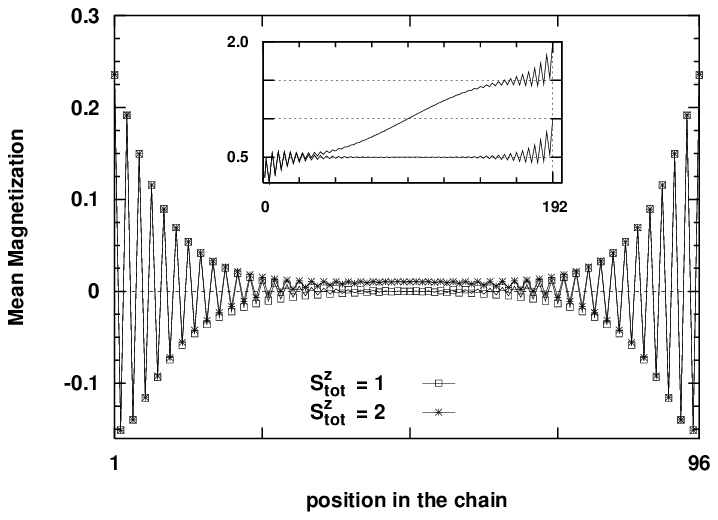}
\includegraphics[scale=1.0]{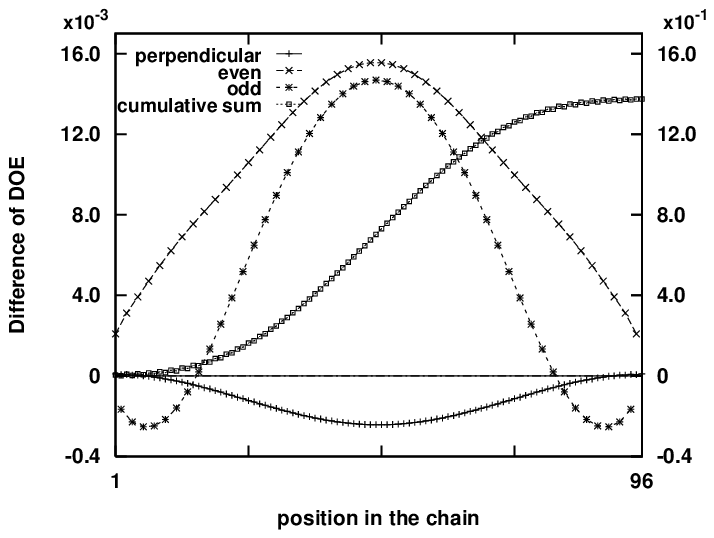}
\caption
{
Computations on a $L=2\times 96$ ladder at the point $(J'/J=-2.5, \gamma=0)$.
 \textit{Up}: Mean magnetization $\langle S^z_i\rangle$ in the states
with $S_{tot}^z=1$ and $S_{tot}^z=2$, as explained in the text.
\textit{Inset}:
Cumulative sum of the magnetization over the whole extent of the ladder and the
 states with $S^z_{\textrm{tot}}=1$ and $S^z_{\textrm{tot}}=2$. The order of
 the sites in the $x$-axis corresponds in
this case to the path used to traverse the ladder in a DMRG sweep.
\textit{Down}: Difference of energy density (DOE) of the excited states
with $S^z_{\textrm{tot}}=1$ and $S^z_{\textrm{tot}}=2$:
$\langle\textbf{S}_{\ell,i}\textbf{S}_{\ell ',i'}\rangle_{S^z=2} -
\langle\textbf{S}_{\ell,i}\textbf{S}_{\ell ',i'}\rangle_{S^z=1}$.
.The scale on the right axis corresponds to the cumulative sum. See text for more explanations.
}
\label{OBCexcitations}
\end{figure}

For generic values of the coupling constants in the Hamiltonian \eqref{2legHam},
this model has been the subject of a series of conjectures based on
exact diagonalization numerical studies \cite{watanabe94} in the absence of
dimerization $\gamma=0$ and analytical
studies using bosonization and NL$\sigma$M mapping \cite{totsuka_suzuki95}
in the presence of dimerization $\gamma\neq 0$. Those numerical methods only
allowed to reach ladder lengths typically of $L=6$ or so, which prevents from
reaching any definitive conclusion on the bulk properties of the system in the
thermodynamic limit. As for the analytical studies, they conjectured the
existence of a possible critical region, but due to the nature of the methods
it is not possible to give its location in terms of the original coupling
constants in the model Hamiltonian \eqref{2legHam}.

\begin{figure}[tb]
\includegraphics[scale=1.0]{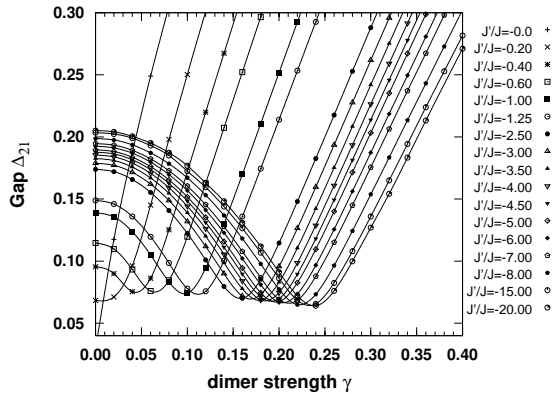}
\caption
{Gap $\Delta_{21}$ computed on a $L=2\times 140$ ladder for different values of
the parameter $J'/J$. Each minimum in the gap belongs to the critical line.}
\label{gap21}
\end{figure}

\section{Critical region}
\label{sect_critical}

One of the main issues in this model \eqref{2legHam} is whether it
exhibits a critical line in the quantum phase diagram of $J'/J$ vs. $\gamma$.
We solve this open problem in the positive by using the DMRG method in finite
version algorithm which provides us with better accuracy values than the infinite
method version, although at the expense of more demanding time computing resources.
The performance of the finite DMRG algorithm is characterized by the following parameters:
the number of states $m$ retained in the truncation process of the RG method, the weight
of the discarded states $w_m$ which is a measure of the DMRG error,
the number of sweeps $n_s$ or iterations of the method after the initial warm-up process and 
 the tolerance $\epsilon$ of the target state energy which controls
the average number of iterations that will need the diagonalization algorithm (Lanczos in our 
case) to compute the target state. 
We shall provide values of these parameters in our numerical computations below.

Before applying the finite-size DMRG method, two important remarks are in order:

\noindent i/ As we shall always work with a fixed value of $L$ the length of the lattice,
the gap $\Delta(J'/J, \gamma)$ is always finite and only in the thermodynamic limit
$L\rightarrow\infty$ it may vanish for certain values of $J'/J$ and  $\gamma$ which define
the critical line we are searching for. Thus, the signature of a gap in $\Delta(J'/J, \gamma)$
for fixed $J'/J$ and varying $\gamma$ will show up as a minimum in the dimerization parameter.
Upon increasing the value of $L$, we shall obtain more robust estimations of the critical value
$\gamma_c(J'/J)$ from the minima $\gamma_{\rm min}(L)$. This is a finite-size scaling analysis
of the DMRG numerical data.

\begin{figure}[htb]
\includegraphics[scale=1.0]{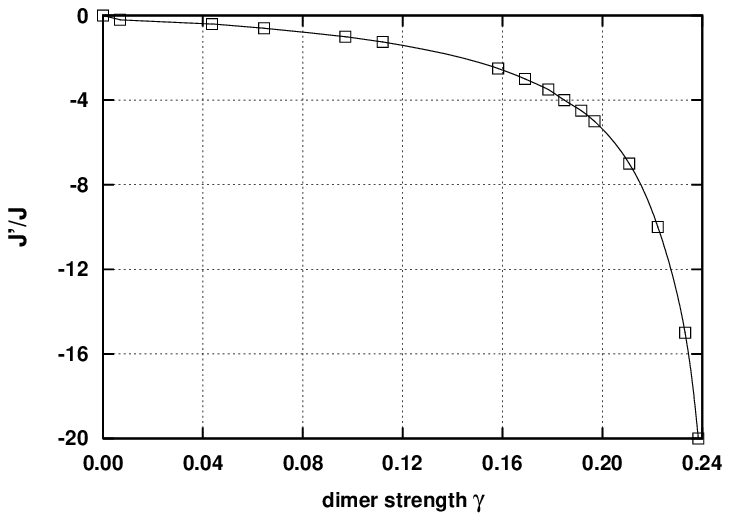}
\includegraphics[scale=1.0]{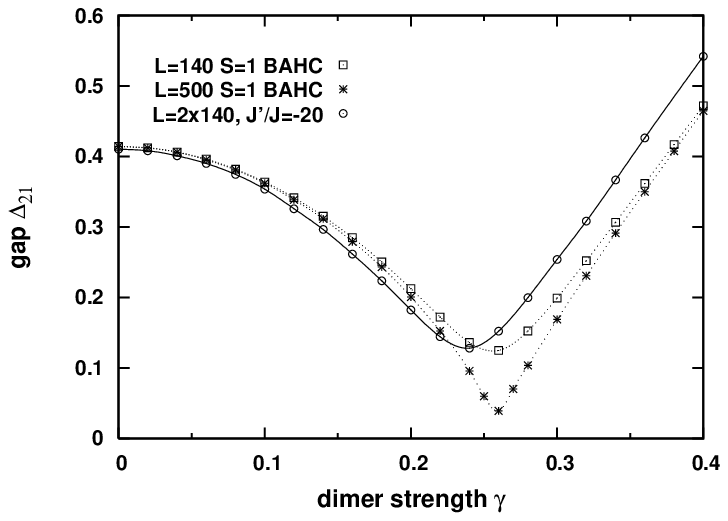}
\caption
{
\textit{Top}:Critical region of a ladder of size $L=2\times 140$.
 The points of the critical line
correspond to the coordinates that minimize $\Delta_{21}(\gamma,J'/J)$. The solid line
is only a guide for the eye.\textit{Bottom}:
The value of the minimum gap for the ladder with $L=2\times 140$ sites is very similar
to the corresponding $S=1$ BAHC with $L=140$ but the value of $\gamma_c$ that minimizes
this gap is still a bit shifted, which constitutes a signal that $J'/J=-20$ is still a low
value to accurately mimic the limit BAHC behaviour. The computations
for the $L=500$ BAHC were performed storing $m=450$ eigenvectors of the density matrix.
}
\label{L500vsL2x140}
\end{figure}

\noindent ii/ The physics of this 2-leg ladder \eqref{2legHam} is richer when the lattice has
open boundary conditions. Moreover, the numerical performance of the DMRG method is also better
under these conditions. However, open boundary conditions must be handled with care in order
to identify the gap $\Delta(J'/J, \gamma)$ we are after.
We shall provide ways to do this identification
by targeting appropriate low-lying states and measuring convenient observables with them.

In particular, using open boundary conditions we have found that the first excited state
lies within the sector with  total z-spin angular momentum $S^z_{\textrm{tot}}=1$,
but in the Haldane phase it converges to the ground state that has $S^z_{\textrm{tot}}=0$
as we take larger sizes of the system. We have then to consider the next excited 
state in the sector
with $S^z_{\textrm{tot}}=2$ to compute the gap of the spectrum as
\begin{equation}
\Delta_{21}:=E_0(S^z_{\textrm{tot}}=2)-E_0(S^z_{\textrm{tot}}=1),
\end{equation}

\begin{figure}[htb]
\includegraphics{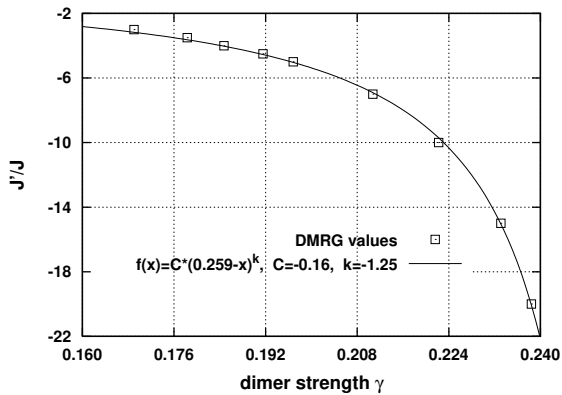}
\caption
{
The region of the critical line in the limit $\vert J'/J\vert \gg 1$ fits very
well to a potential function of
the form $J'/J=C(0.259-\gamma_c)^k$, with $C=-0.16\pm0.01$ and $k=-1.25\pm0.01$.
}
\label{fit}
\end{figure}

The reason for considering $\Delta_{21}$ instead of $\Delta_{10}:=
E_0(S^z_{\textrm{tot}}=1)-E_0(S^z_{\textrm{tot}}=0)$ as the gap of
the system, can be justified as follows: in the complete dimerized
limit $\gamma=1$, it is clear that the difference in energy
between two arbitrary consecutive levels is the same, and
corresponds exactly to the energy needed to promote one
singlet bond to a triplet. The argument for the limit $\gamma=0$
makes use of the properties of Haldane phases, where it is known
to appear a non bulk excitation due to the existence of virtual
spins at the end of the chain. Our conclusion is that the lowest
lying state with $S_z^{\textrm{tot}}=2$ consists on a
superposition of two kind of excitations, namely the Haldane
non-bulk triplet mentioned before and the bulk itself, also giving
a triplet. Considering this scheme, in order to obtain the gap
related to the bulk excitations we have to substract the non bulk
excitations present in the lowest lying states of sectors
$S_z^{\textrm{tot}}=1$ and $S_z^{\textrm{tot}}=2$.

 In Fig.\ref{OBCexcitations} we show rigourous comparations of these two states in the Haldane limit
$\gamma=0$.
Computations have been done on a $L=2\times 96$ ladder at the point $(J'/J=-2.5, \gamma=0)$.
On the \textit{up} part of the figure, we plot the  mean magnetization $\langle S^z_i\rangle$
in the states with $S_{tot}^z=1$ and $S_{tot}^z=2$, computed in one leg of the
ladder, since due to the
symmetry of the Hamiltonian,  the magnetization is the same in both legs. As a check of the
accuracy of our computations we observed that the results
in both legs are the same up to the fifth or sixth decimal digit.
In the \textit{Inset} of that figure, we show the
cumulative sum of the magnetization over the whole length of the ladder and the
 states with $S^z_{\textrm{tot}}=1$ and $S^z_{\textrm{tot}}=2$. The order of
 the sites in the $x$-axis corresponds in
this case to the path used to traverse the ladder in a DMRG sweep.
In the \textit{down} part of this figure, we plot the difference of energy density of the excited states
with $S^z_{\textrm{tot}}=1$ and $S^z_{\textrm{tot}}=2$:
$\langle\textbf{S}_{\ell,i}\textbf{S}_{\ell ',i'}\rangle_{S^z=2} -
\langle\textbf{S}_{\ell,i}\textbf{S}_{\ell ',i'}\rangle_{S^z=1}$
 The difference has been divided into three contributions: the contribution labelled
with \textit{even} stands for links involving sites in the same leg and the even sublattice
($\ell '=\ell, i=2k, i'=2k+1$), \textit{odd} involves links joining sites in the
same leg and the odd sublattice ($\ell '=\ell, i=2k-1, i'=2k$), and \textit{perpendicular}
denotes links among legs ($\ell=1, \ell '=2, i=k, i'=k$).
The cumulative sum of the difference of the various contribution, measured
in the right axis scale, is also shown.
Interestingly enough, we can observe the magnetization pattern at the ends being almost
 identical in the states with $S^z_{tot}=1$
 and $S^z_{tot}=2$. The contribution to the $z$-axis projection
of the spin coming from the ends is equal to 1 in both cases. Notice also
 that the difference of the
density of energy between these states is close to zero at the ends, while it becomes clearly
apreciable in the bulk. All these facts strongly support the picture of a non-bulk  triplet
 excitation with the same nature in both states, that leaves the bulk of the
 chain with a neat value of the projection equal to
$S^z_{\small{\textrm{bulk}}}=0$ and $S^z_{\small{\textrm{bulk}}}=1$ and gives a strong hint on the equivalence
of $\Delta_{10}$ with periodic boundary conditions and $\Delta_{21}$ for open systems.


\begin{figure}[tb]
  \includegraphics[scale=1.0]{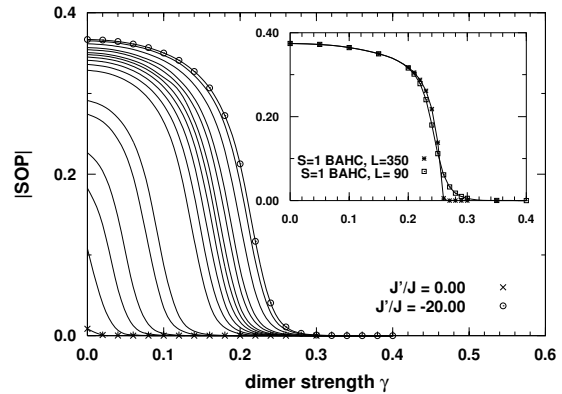}
\caption
{
String order parameter computed for a ladder consisting on $L=2\times96$ sites.
 The SOP has
been computed forming the triplets with adjacent $S=1/2$ spins located in different legs.
Clearly the value of this parameter is non-vanishing in the low region of $\gamma$, where
the system is in the Haldane phase. The dimer phase is nonetheless characterized by
a vanishing SOP.\textit{Inset}: SOP computed for a $S=1$ BAHC. The resemblance between
both systems is evident in the region with $\vert J'/J\vert \gg 1$.
}
\label{SOP}
\end{figure}
\begin{figure}[tb]
\includegraphics[scale=0.5]{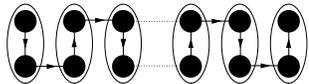}
\caption{A picture of the path used to compute the string order parameter in the
2-leg ladder with columnar bond alternation. The ellipses mean that the sites within them
are forming a triplet.}
\label{ladderSOP}
\end{figure}

After this previous analysis to identify the states needed to target the gap of the system,
we present in Fig.\ref{gap21}  some values of the gap $\Delta_{21}$ for a ladder
consisting on $L=2\times 140$ sites, at different regions of the parameter
space. Computations have been performed retaining $m=300$ states of the density matrix
and the grid used to explore the phase diagram is $\gamma \in [0,0.4]$,
and $-J'/J=\{0.00, 0.20, 0.40, 0.60, 1.25, 2.50, 3.00 ,3.50 ,4.00, 4.50, 5.00,\linebreak
6.00, 7.00, 10.00, 15.00, 20.00\}$. The existence of a set of minima in the
function $\Delta_{21}(\gamma,J'/J)$ is clear in this graph, although they shall become more
distinguishable  as we move to higher values of  $\vert J'/J\vert$.

As an instance of the accuracy of our results, we point out that a
systematic examination of the error in each of the truncations of our
DMRG computations reveals
that the highest values in the whole process are of the order of $w_m \sim 10^{-8}$, and mostly
they are of order $w_m \sim 10^{-10}$. To obtain a suitable acuracy in the results we have 
 set the number of sweeps $n_s=2$ and the tolerance to $1e-9$. To compute with enough precision the critical value
$\gamma_c(J'/J)$ that minimizes the gap it becomes necessary to use large amount of data.
On this regard, we have used interpolated values resulting from the DMRG computations.

\begin{figure}[tb]
\includegraphics[scale=1.0]{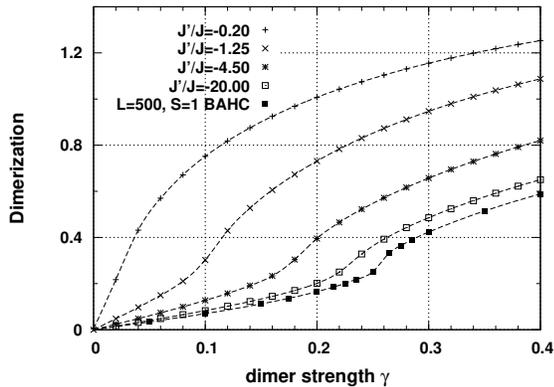}
\caption
{
Different values of the dimerization parameter
$D_i:=\sum_{\ell=1,2}\langle \textbf{S}_{\ell,i-1}\textbf{S}_{\ell,i}\rangle -
\langle \textbf{S}_{\ell,i}\textbf{S}_{\ell,i+1}\rangle$ computed in the middle of a ladder
of $L=2\times96$ sites.
Since we are explicitly
introducing some staggering in the
Hamiltonian \eqref{2legHam}, the dimerization parameter is non-vanishing even in the Haldane phase. However,
the shape of the graphs seem to have an inflexion at the critical point.
The graph corresponding to the $S=1$ BAHC, has been scaled down by a factor $1/2$ due
to the effective coupling constant of the ladder, which is known to be half the constant
corresponding to the BAHC.
}
\label{dimerization}
\end{figure}

Now, we can detect the presence of a critical line in the quantum phase diagram
separating gapped phases. In Fig.\ref{L500vsL2x140} we plot the critical region consisting of the
coordinates for each minimum in Fig.\ref{gap21}. In earlier
studies \cite{Salerno2006}, we  placed the critical point of the $S=1$ bond-alternating Heisenberg
chain (BAHC) at $\gamma_c=0.259$. The curve shown
in Fig.\ref{L500vsL2x140} shows a vertical asymptota that is still a bit off from this limiting value
corresponding to the region $\vert J'/J\vert \gg 1$,  but this is simply because we have chosen
a value of $J'/J = -20$ which is still not big enough and also due to finite-size effects on the
2-leg ladder.
In the lower plots of Fig.\ref{L500vsL2x140} we address these possibilities  by comparing
our ladder in the strong ferromagnetic limit with a pure $S=1$ BAHCs with different sizes.
Two parameters are important in this discussion, namely, the value $\gamma_c(J'/J)$ that minimizes the gap, and
the value of the gap itself at this point $\Delta_{21}(\gamma_c)$. As we can observe in
Fig.\ref{gap21}, the value of $\Delta_{21}(\gamma_c)$ does not strongly depend on the
particular choice of the coupling constant ratio $J'/J$, while it is definitely influenced by the size of
the system. In Fig.\ref{L500vsL2x140}(lower) it is shown that the shift of $\gamma_c$ computed for
two $S=1$ BAHCs with different sizes, but still large enough both, is less noticeable than
the difference in their value of $\Delta_{21}(\gamma_c)$. It is clear the similarity of
this magnitude in the case of the ladder and the corresponding BAHC, as well as the shift
in the value of $\gamma_c$.
All this make us conclude that in order to attain a better convergence with
the $S=1$ BAHC and a better estimate of the critical asymptota $\gamma_c=0.259$, we shall
increase the strength of the ferromagnetic coupling rather than the size of the system.

As we have a set of numerical data from the finite-size analysis of the critical line,
we can also make a numerical estimation of the criticality curve.
In fig. \ref{fit} we present a fit of the
 critical curve in the region close to $\gamma_c\simeq0.259$.
We choose as trial function for this fitting an inverse power law with some
coefficients and exponents that are fixed by our numerics, namely,
\begin{equation}
J'/J = \frac{C}{(0.259-\gamma)^k}.
\end{equation}
The fitting yields the following estimations for the values of
the parameters $C$ and $k$ that best fit the data:
$C=-0.16\pm0.01$ and $k=1.25\pm0.01$, and for simplicity the value of $\gamma_c=0.259$
is taken for fixed.

\begin{figure}[tb]
\includegraphics[scale=.650]{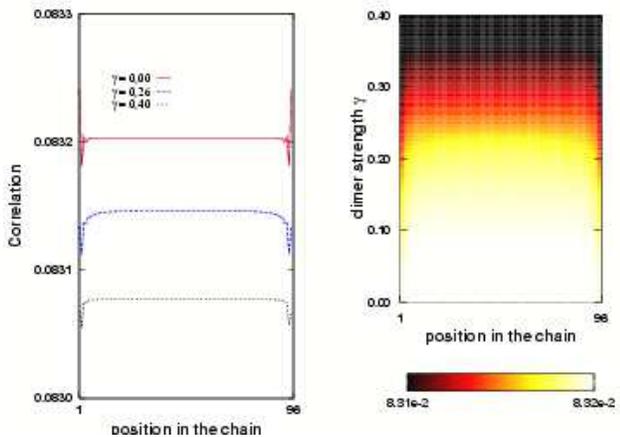}
\caption
{
Correlation $\langle S^z_i(1)S^z_i(2)\rangle$ in the ground state of a $L=2\times 96$
sites ladder and $J'/J=-20$. The plot on the left shows the correlation value for some
arbitrary values of $\gamma$. The plot on the right shows the value of this
magnitude in the whole region of the parameter $\gamma$.
}
\label{colormap0}
\end{figure}

\section{Haldane and Dimer phases}
\label{sect_haldanedimer}

Once we have established the existence of a critical line
in the quantum phase diagram of the model \eqref{2legHam}, it is
natural to wonder about the two gapped phases that this line
separates. More specifically, whether they are different or not
and their identification as quantum phases in the framework
of strongly correlated systems.

The possible nature of those phases can be guessed from
the strong ferromagnetic limit $1\ll\vert J'/J\vert$ of the ladder, effectively leading to
the $S=1$ BAHC. The phases of this chain are known to be the massive Haldane phase,
separated by a critical point from the also massive dimer phase.
 To test the nature of each phase, we will resort to two different order parameters.
The Haldane phase is known to exhibit a particular hidden order that can be measured by
the string order parameter (SOP) \cite{sop1}, \cite{sop2}.
The definition of this operator for a spin-1 chain is as follows:
\begin{equation}
O(\ell)=\langle S^z_1\prod_{k=2}^{\ell-1} e^{i\pi S^z_k}S^z_\ell \rangle
\end{equation}
This operator acting on our ground state measures how far it is from a spin liquid
 N\'eel state consisting on a sequence of $S=1$ spins such that every
spin with projection $S_i^z=\pm 1$ is followed by $S_{i+k}^z=\mp 1$ and $S_{i+k'}^z=0$
for every $0<k'<k$.

When we deal with $S=1/2$ particles, to compute the SOP we have to define the pairs of
particles which are most likely to couple to give a triplet and compute the SOP along the path
connecting them.
In our case, the existence of a ferromagnetic coupling clearly suggests that the triplets
will result via this coupling. It is also worth recalling that the SOP is a parameter suited to
work with translational invariant systems. In order to correctly estimate the SOP in open
systems, we must restrict the computation to a region shorter than the whole length of
the chain where end-effects are negligible and only bulk physics is relevant.
In Fig.\ref{SOP} we show the SOP computed traversing the
path shown in Fig.\ref{ladderSOP}. We can observe a non-vanishing SOP in the Haldane region,
while it rapidly decays to zero in the dimer phase. The inset shows the SOP computed for
a $S=1$ BAHC and the resemblance between both systems is apparent.

Therefore, the phase below the critical line in the numerical phase diagram of Fig.\ref{L500vsL2x140}
is a gapped Haldane phase.

\begin{figure}[tb]
\includegraphics[scale=.650]{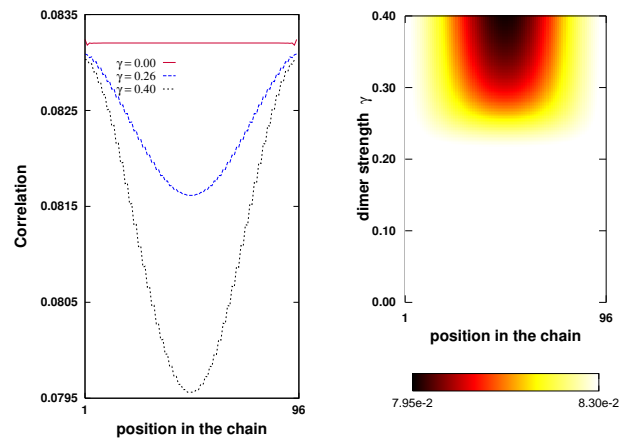}
\caption { Correlation $\langle S^z_i(1)S^z_i(2)\rangle$ in the
first lying excited state in the sector with
$S^z_{\textrm{tot}}=1$ of a $L=2\times 96$ ladder and $J'/J=-20$.
}
\label{colormap1}
\end{figure}

\begin{figure}[tb]
\includegraphics[scale=.70]{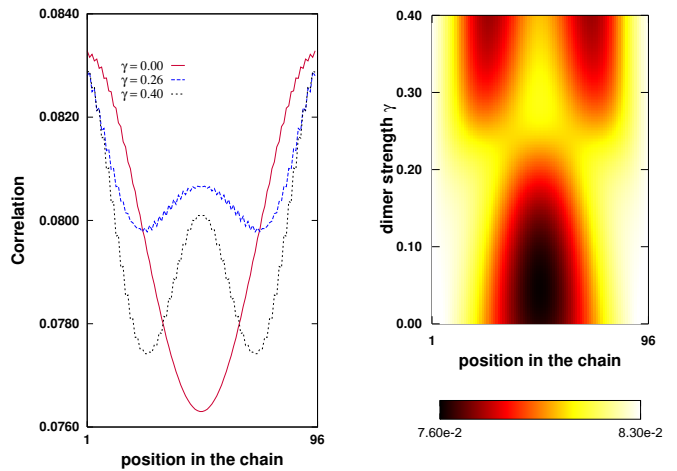}
\caption
{
Correlation $\langle S^z_i(1)S^z_i(2)\rangle$ in the first lying excited state
in the sector with $S^z_{\textrm{tot}}=2$ of a ladder consisting on $L=2\times96$ sites
and $J'/J=-20$.
}
\label{colormap2}
\end{figure}

As for the region above the critical line in Fig.\ref{L500vsL2x140}, we have guessed from the strong coupling limit
that it may be a dimer phase.
The structure of a dimer phase is such that full translational invariance symmetry
 of the system is broken by one unit cell of the lattice.
This situation can be detected by means of the dimerization parameter, which can be defined
for our particular 2-leg ladder as
\begin{equation}
D_i:=
\sum_{\ell=1,2}\langle \textbf{S}_{i-1}(\ell)\cdot \textbf{S}_{i}(\ell)\rangle -
\langle \textbf{S}_{i}(\ell)\cdot \textbf{S}_{i+1}(\ell)\rangle.
\end{equation}
\begin{figure}[tb]
\includegraphics[scale=1.0]{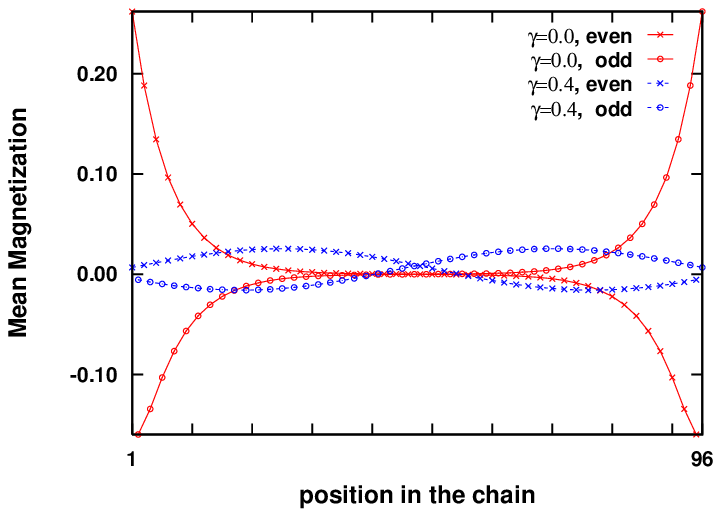}
\includegraphics[scale=1.0]{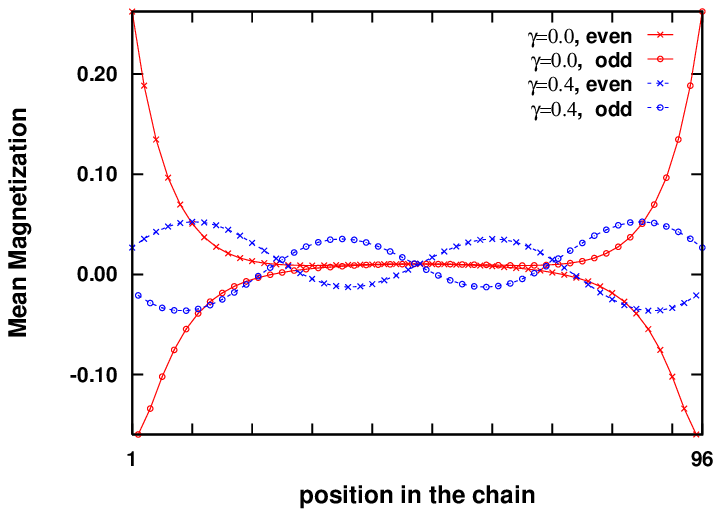}
\caption
{
Magnetization curves in a $L=2\times 96$ ladder for the lowest lying states in the sectors
$S^z_{\textrm{tot}}=1$ (\textit{top}), and $S^z_{\textrm{tot}}=2$ (\textit{bottom}).
The curves are separated in the different sublattices consisting on the sites ocuppying
odd or even positions. Notice that both states present a peaked magnetization at the ends
for $\gamma=0.0$ well into the Haldane phase, while it vanishes in the dimer phase with
 $\gamma=0.4$.
}
\label{Szgraph}
\end{figure}
The subindex $i$ is necessary since open systems are intrinsically not translationally
 invariant. In Fig.\ref{dimerization}
it is plotted the dimerization parameter of the ladder measured in the middle of the chain.
Since staggering is explicitly introduced into the Hamiltonian \eqref{2legHam}, the order parameter vanishes
only at $\gamma=0$, but is finite even in the Haldane phase. Nevertheless, our plots clearly
exhibit different behaviours related to the convexity of the parameter at each phase.
This observation indicates that an accurate estimation of the point of inflexion in
the dimerization parameter could be used as a measure of the critical point separating both phases.

We have also performed some measurements in the ladder to give more hints to
understand the nature of both phases. Figures \ref{colormap0},
\ref{colormap1}, and \ref{colormap2} show the correlation
$\langle \textbf{S}_i(1)\textbf{S}_i(2) \rangle$ between sites in the perpendicular rungs.
The pattern of the correlation can be understood by noticing that the correlation between two
isolated $S=1/2$ spins coupled to give a singlet is
$\langle \textbf{S}_1\textbf{S}_2\rangle/3=-1/4$ while it equals $1/12$ if the spins form a
triplet. From these values we observe that the perpendicular rungs in the ground state are
forming triplets and the distribution is uniform all along the ladder. In the excited states
however, the triplet strength of some rungs is weakened, signaling the presence of magnon-like
excitations, also apparent in Fig.\ref{Szgraph}. The nature of the non-bulk excitation
present in the Haldane phase is not magnon-like and that explains the different number of kinks
in the Haldane and dimer phase in Fig.\ref{colormap1} and Fig.\ref{colormap2}.

\section{Conclusions}
\label{sect_conclusions}

We have determined the existence of a critical line in the quantum phase
diagram of a 2-leg ladder with columnar dimerization and ferromagnetic vs.
antiferromagnetic couplings. In this study, we use the finite-size system
DMRG method which allows us to give the location of that critical curve.
Moreover, we have clearly identified the two phases separated by the critical
line to be a Haldane phase and a dimer phase. This identification is carried out
by measuring the string order parameter and the dimerization order parameter in the
whole range of values of the coupling constant ratio $J'/J$ and dimerization parameter $\gamma$.

As a byproduct, we have introduced a systematic analysis of the
spins at the borders of the open 2-leg ladder lattice.
Our model is based on $S=\half$ spins, then these end-chain spins
exhibit physical effects of their own. They are real
spins unlike the virtual spins appearing in integer spin chains or ladders.
Their physics is specially interesting when the system size is finite,
and even during the process of reaching the thermodynamic limit they produce non-trivial finite-size effects
along the way. These facts difficult the technical analysis of the opening or closing of a gap
in the low-lying  spectrum of a 2-leg ladder with open boundary conditions.
We have solved these difficulties by analyzing the ground state and low-lying energy excitations
 with respect to their bulk and boundary properties such as local magnetization and the like.
 With this information, it is possible to identify which states contribute to the gap in thermodynamic limit.
 These low-lying states have a definite total spin $S^z$ and they can be targeted with the DMRG method.
In this fashion, we have been able to identify the gapped or gapless behaviour of the model within the framework
of the DMRG with open boundary conditions.

 \noindent {\em Acknowledgements}
 We acknowledge financial support from  DGS grants  under contracts BFM 2003-05316-C02-01, 
FIS2006-04885, and the ESF Science Programme INSTANS 2005-2010.

\end{document}